\documentclass[11pt]{article}
\usepackage[colorlinks=true,linkcolor=magenta,citecolor=magenta]{hyperref}
\usepackage[letterpaper, margin=1in]{geometry}

\title{Hyper-Reduced Autoencoders for Efficient and Accurate \\ Nonlinear Model Reductions}

\usepackage[utf8]{inputenc}
\usepackage[T1]{fontenc}
\usepackage{authblk}
\usepackage[numbers]{natbib}

\usepackage{authblk} 

\author[1]{Jorio Cocola}
\author[2]{John Tencer}
\author[3]{Francesco Rizzi}
\author[2]{Eric Parish}
\author[2]{Patrick Blonigan}

\affil[1]{Harvard University}
\affil[2]{Sandia National Laboratories}
\affil[3]{NexGen Analytics}

\usepackage{enumerate}
\usepackage{comment}
\usepackage{amsfonts,amsmath,bbm,amssymb, amsthm,amssymb, upgreek,mathtools}
\usepackage[linesnumbered,ruled,vlined]{algorithm2e}
\usepackage{float}
\usepackage{bm}

\usepackage[toc,page,header]{appendix}
\usepackage{minitoc}

\usepackage{amsfonts,amsmath,bbm,amssymb, amsthm,amssymb, upgreek,mathtools}
\usepackage{wrapfig}
\usepackage{comment}

\usepackage[english]{babel}
\usepackage{graphicx}
\usepackage{float, amssymb,amsmath}
\usepackage{ upgreek }
\usepackage{tcolorbox, color}
\usepackage{graphicx}
\usepackage{xcolor}
\usepackage{import}
\usepackage{paracol}

\usepackage{empheq}

\usepackage{thmtools}

\newtheorem*{thm*}{Theorem}

\usepackage{varwidth}

\usepackage{tcolorbox}

\usepackage{subcaption}

\usepackage{amssymb,amsthm,amsmath,amssymb,bbold,bm}

\def\relu{{\textrm{ReLU}}}
\def\leakyRelu{{\textrm{LeakyReLU}}}

\def\vmu{{\bm{\mu}}}

\def\vb{{\bm{b}}}

\def\ve{{\bm{e}}}
\def\vf{{\bm{f}}}
\def\vg{{\bm{g}}}
\def\vh{{\bm{h}}}

\def\vp{{\bm{p}}}

\def\vr{{\bm{r}}}

\def\vu{{\bm{u}}}

\def\vx{{\bm{x}}}
\def\vy{{\bm{y}}}
\def\vz{{\bm{z}}}

\def\vmu{{\bm{\mu}}}

\def\mA{{\bm{A}}}

\def\mI{{\bm{I}}}
\def\mJ{{\bm{J}}}

\def\mP{{\bm{P}}}
\def\mQ{{\bm{Q}}}
\def\mR{{\bm{R}}}

\def\mU{{\bm{U}}}

\def\mW{{\bm{W}}}
\def\mX{{\bm{X}}}
\def\mY{{\bm{Y}}}

\DeclareMathAlphabet{\mathsfit}{\encodingdefault}{\sfdefault}{m}{sl}
\SetMathAlphabet{\mathsfit}{bold}{\encodingdefault}{\sfdefault}{bx}{n}

\newcommand{\R}{\mathbb{R}}

\newcommand{\dd}{\mathrm{d}}
\DeclareMathOperator*{\argmin}{arg\,min}

\usepackage{bm}
\def\mPhi{{\bm{\Phi}}}

\begin{document}

\doparttoc 
\faketableofcontents 

\maketitle

\begin{abstract}
	Projection-based model order reduction on nonlinear manifolds has been recently proposed for problems with slowly decaying Kolmogorov n-width such as advection-dominated ones. These methods often use neural networks for manifold learning and showcase improved accuracy over traditional linear subspace-reduced order models. A disadvantage of the previously proposed methods is the potential high computational costs of training the networks on high-fidelity solution snapshots. In this work, we propose and analyze a novel method that overcomes this disadvantage by training a neural network only on subsampled versions of the high-fidelity solution snapshots. This method coupled with collocation-based hyper-reduction and Gappy-POD allows for efficient and accurate surrogate models. We demonstrate the validity of our approach on a 2d Burgers problem.
\end{abstract}

\section{Introduction}
   	\subsection{Projection-based Reduced Order Models}


We consider high dimensional parameterized dynamical systems resulting from the spatial discretization of  Partial Differential Equations (PDEs). In particular, Full Order Models (FOMs) of the form:
\begin{equation}\label{eq:FOM}
    \frac{\dd \vy}{\dd t} = \vf(\vy, t; \vmu), \qquad \vy(0; \vu) = \vy_0(\vmu)
\end{equation}
where $t \in [0,T]$ is the time variable and $T>0$ the final time, $\vmu \in \R^{n_\mu}$ denotes system parameters, $\vy = \vy(t;\vmu) \in \R^N$ is the state, $\vy_0$ is the parameterized initial condition and $\vf:\R^N \times \R_+ \times \R^{N_\mu} \to \R^{N}$ denotes the nonlinear velocity. 
In particular we are interested in the case where the dimension $N$ of the FOM \eqref{eq:FOM} is much larger than the dimension $n_\mu + 1$ of the solution manifold \[
\mathcal{M} = \{(\vy(t, \mu) \: | \: t \in [0,T], \,\vmu \in \R^{n_\mu} \}.
\] 

We study projection-based reduced order models for the FOM \eqref{eq:FOM} based on a nonlinear approximation of the solution manifold. These methods consider approximations of the solutions $\vy \in \mathcal{M}$ of the form 
\begin{equation}\label{eq:nonLinDec}
    \vy \approx \tilde{\vy} := \vg(z)
\end{equation}
where $\vg: \R^{k} \to \R^N$ is a linear or nonlinear map, often termed \textit{decoder}, and $\vz \in \R^{k}$ is the \textit{reduced order state} or \textit{latent vector}. Note that the $\vg$ generates the nonlinear manifold  
\[
\mathcal{G} = \{ \vg(\vz) \,|\, \vz \in \R^k  \}
\]  
approximation of the solution manifold $\mathcal{M}$.


\subsection{Manifold LSPG ROM}

To obtain the Reduced Order Model (ROM) corresponding to the FOM \eqref{eq:FOM}, and the approximations \eqref{eq:nonLinDec}, we use manifold Least-Squares Petrov-Galerkin Projection (LSPG) \cite{lee2020model}.
This method consists in first applying a time-discretization method to the FOM \eqref{eq:FOM}, and then projecting the resulting residual on the trial manifold identified by the maps \eqref{eq:nonLinDec}.

A temporal discretization of \eqref{eq:FOM} on the time grid $t^0, \dots, t^{N_t}$  yields the FOM O$\Delta$E system 
\begin{equation}\label{eq:ODeltaE}
	\vr^n(\vy^n; \vy^{n-1}, \dots, \vy^{n-\ell}, \vmu) = 0 \quad\text{for}\;\; n = 1, \dots, N_t, 
\end{equation}
where $\vr^n: \R^{N\cdot(\ell+1)} \times \R^{N_\mu} \to \R^N$ is the residual, $\vy^n$ is the state at the $n$-th timestep and $\ell = \ell(n)$ is the width of the stencil of the discretization at time $t^n$. 
 

Substituting then the approximation $\vy_n \leftarrow \tilde{\vy}_n$ defined in \eqref{eq:nonLinDec} and  minimizing the residual \eqref{eq:ODeltaE}, we obtain the time discrete ROM:
\begin{equation}\label{eq:projectODeltaE}
    \vz^n := \argmin_{\vz \in \R^k} \| \vr^n(\vg(\vz); \tilde{\vy}^{n-1}, \dots, \tilde{\vy}^{n-\ell(t^n)}, \vmu) \|_2^2 \quad \text{for}\;\; n = 1, \dots, N_t
\end{equation}
where $\tilde{\vy}^{n} = \vg(\vz^n)$ provides an approximation of the state $\vy^n$.
\subsection{Collocation-based hyper-reduction}
Even though the ROM states $\{\vz^n\}_n$ have small dimensions $k \ll N$, the residual $\vr$ in \eqref{eq:projectODeltaE} still scales like the dimension $N$ of the FOM and solving \eqref{eq:projectODeltaE} would incur in cost that also scales with $N$. 
Hyper-reduction methods attempt to alleviate this issue by constructing cheap approximations of the residual \eqref{eq:projectODeltaE} \cite{ryckelynck2005priori, farhat2014dimensional, lauzon2022s}.
In particular, \textit{collocation-based hyper-reduction} samples the nonlinear residual \eqref{eq:ODeltaE} at a prescribed set of indices referred to as the \textit{sample mesh}. Let $\mP_c \in \{0,1\}^{n_c \times N}$ be the sampling matrix that picks out the $n_c$ indices of the residual $\vr^n$, then the manifold LSPG \eqref{eq:ODeltaE} with collocation takes the form:
\begin{equation}\label{eq:hyperODeltaE}
    \vz^n := \argmin_{\vz \in \R^k} \| \mP_c \vr^n(\vg(\vz); \tilde{\vy}^{n-1}, \dots, \tilde{\vy}^{n-\ell(t^n)}, \vmu) \|_2^2 \quad \text{for}\;\; n = 1, \dots, N_t.
\end{equation}

In many cases of interest, because of the locality of the operators of the original PDE from which the FOM \eqref{eq:FOM} is derived and the locality of the space-discretization used, the states $\{ \tilde{\vy}^{j} \}_{n - \ell(t^n)}^n$ need to be evaluated only on a subset of indices. We refer to this set of indices as \textit{stencil mesh} and observe that in general its cardinality $n_s$ satisfies $n_c < n_s < N$.

Notice that in practice it is not necessary to form the full residual $\vr$ and the states $\{ \vy^j \}_{j=n-\ell}^n$ in \eqref{eq:hyperODeltaE}, as it suffices to keep track of $\vr$ on the sample mesh and of $\{\vy^j\}_{j=n-\ell}^n$ on the stencil mesh.

   	\subsection{Our contributions}

The nonlinear map $g: \R^{k} \to \R^{N}$ used to reduce the FOM \eqref{eq:FOM} and approximate the solution manifold $\mathcal{M}$ is, in general, unknown and learned from data. 
Recent works propose to parameterize and learn $g$ using neural networks \cite{lee2020model,choi,romor2022non}, demonstrating improved performances (more accurate predictions) with respect to traditional approaches such as Proper Orthogonal Decomposition (POD). 
One disadvantage of these approaches is that the training of the neural networks models is often computationally costly requiring many passes of Stochastic Gradient Descent over the data with thousand of computations of the order $\mathcal{O}(N)$, which, for large scale problems can be impractical. 

In this work, we develop and analyze a method to overcome the above issue by learning a nonlinear map $\hat{\vg}$ which predicts the values of the solutions $\vy$ only on the stencil points. 
Specifically, let $\mP_s \in \{0,1\}^{n_s \times N}$ be the sampling matrix that picks out the $n_s$ indices of the stencil mesh, our goal will then be to approximate the solutions $\vy \in \mathcal{M}$ restricted on the stencil mesh
\begin{equation}\label{eq:ghat}
    \mP_s \vy \approx \hat{\vy} := \hat{\vg}(\vz),
\end{equation}
where $\hat{\vg} : \R^k \to \R^{N_s}$ is a nonlinear map that we refer to as the \textit{``hyper-decoder''}. 
This approach constitutes a departure from the previous nonlinear projection-based model reduction methods, in that the primary objective of the nonlinear map $\hat{g}$ is to approximate the reduced manifold 
\[
\mathcal{M}_s = \{\mP_s^T \vy(t, \vmu) \: | \: t \in [0,T], \,\vmu \in \R^{n_\mu} \},
\] 
rather than directly the full manifold $\mathcal{M}$. 

We can then use our proposed  ``hyper-decoder'' for solving the manifold LSPG with collocation which, with some abuse of notation, takes the form
\begin{equation}\label{eq:hyperODeltaE_hat}
    \vz^n := \argmin_{\vz \in \R^k} \| \mP_c \vr^n(\hat{\vg}(\vz); \hat{\vy}^{n-1}, \dots, \hat{\vy}^{n-\ell(t^n)}, \vmu) \|_2^2 \quad \text{for}\;\; n = 1, \dots, N_t,
\end{equation}
where $\hat{\vy}^{n} = \hat{g}(\vz^n)$ provides an approximation of the  state $\mP_s^T \vy^n$ restricted on the stencil mesh.

Once the approximation $\hat{\vy}^n$ is obtained, we estimate the full state $\vy^n$ by employing Gappy-POD \cite{everson1995karhunen}.

In summary, our method uses recent advances in deep learning to learn an accurate representation of the nonlinear manifold of the subsampled solutions. 
This nonlinear representation is used to define the reduced order models, which are then solved via LSPG. 
We then use more traditional and fast methods based on efficient matrix factorization to obtain accurate predictions of the high-fidelity states. 
In the next section, we detail our proposed approach.

%


%
\section{Hyper-reduced Decoder}\label{sec:hyp_dec}
   	In this section, we detail our nonlinear model reduction method based on the hyper-reduced decoder. 

The first step in our proposed method is the selection of the indices of the sample mesh and the construction of the stencil mesh. This step is described in Section \ref{subsec:stencil_mesh}. In Section \ref{subsec:AE} we describe how once a set of stencil mesh indices is fixed we can train a reduced decoder $\hat{g}$ using high-fidelity solutions of \eqref{eq:FOM} restricted on the stencil mesh. The reduced decoder $\hat{g}$ can then be used in the reduced order model \eqref{eq:hyperODeltaE} to estimate the restriction of the FOM states $\vy$ on the stencil mesh (as in \eqref{eq:ghat}). In Section \ref{subsec:gappyPOD} we then recall the Gappy-POD method and how it can be used to efficiently recover ${\vy}$ from $\hat{\vy}$.

In this section, we will assume that \eqref{eq:FOM} has been solved for $n_\text{train}$ number of parameters $\mu \in \R^{N_\mu}$ and for $N_t$ number of time steps. The high-fidelity solutions have then been collected in a snapshots matrix
\[
	\mY_{\text{train}} = \big[ \vy^0_{\mu_1}, \dots, \vy^{N_t}_{\mu_1}, \dots,  \vy^{0}_{\mu_{n_\text{train}}}  , \dots,  \vy^{N_t}_{n_\text{train}}\big] \in \R^{N \times n_\text{train}\,(N_t + 1)}.
\]
We will denote by $\mPhi \in R^{N \times n_\text{train}\,(N_t + 1)}$ the orthogonal (POD) matrix obtained the Singular Value Decomposition (SVD) of the snapshot matrix $\mY_{text{train}}$. We also use $\mathcal{Y}_\text{train}$ to denote the set of  high-fidelity solutions $\mathcal{Y}_\text{train} = \{ \vy^0_{\mu_1}, \dots,  \vy^{N_t}_{\mu_1}, \dots,  \vy^{0}_{\mu_{n_\text{train}}}  , \dots,  \vy^{N_t}_{n_\text{train}}\}$

   	\subsection{Sample and stencil mesh generation}\label{subsec:stencil_mesh}


The sample mesh indices are selected to allow the efficient and accurate approximation of the minimization problems \eqref{eq:ODeltaE} via the reduced minimization problem \eqref{eq:hyperODeltaE}. 
In GNAT \cite{carlberg2013gnat}, for example, a snapshot matrix of residuals together with a greedy method is used to select the sample indices. In this work, motivated by the recent GNAT-SNS  \cite{choi2020sns}, we use instead the solution snapshots $\mX$ with a pivoted QR method \cite{berry2005algorithm} of the POD matrix. This leads to reduced storage requirements and a reduced number of large-scale SVD performed. Algorithm \ref{algo:QRSampling} details the steps in our proposed sampling strategy (in Python notation).  

	Once the indices of the sample mesh are obtained, we generate the indices of the stencil mesh considering those indices that are needed to evaluate the residual on the sample mesh. These indices will be adjacent to the indices on the sample mesh and will be dictated by the spatial discretization used to discretize the original PDE from which \eqref{eq:FOM} is derived.
	%

\begin{algorithm}
\DontPrintSemicolon
\KwIn{Number of sample mesh indices $n_c$, POD matrix $\mPhi$.}
\KwOut{{Sample mesh indices $\mathcal{I}_{n_c}$.}}
	${\mPhi_{n_c}} \gets \mPhi[:,:n_c]$. \\
    Compute the QR decomposition of $\mPhi_{n_c}^T$ with column pivoting: $\mPhi_{n_c}^T = \mQ \mR \mP^T$. \\
    Let $\vp\in \R^N$ the vector such that $\mP = \mI_N[:,\vp]$.\\
\Return{$\vp[:n_c]$.}
\caption{{\sc Sample mesh indices selection via pivoted QR method} }
\label{algo:QRSampling}
\end{algorithm}
   	\subsection{Non-Linear Manifold Learning via Noisy Auto-Encoders}\label{subsec:AE}

The next step in our proposed strategy is to learn the map $\hat{\vg}: \R^{k} \to \R^{n_s}$ in \eqref{eq:ghat}. Notice that for the purpose of obtaining accurate reduced models of the FOM \eqref{eq:FOM}, the range of the map $\hat{\vg}$ needs only to approximate a \textit{superset} of the reduced manifold $\mathcal{M}_s$. This in particular means that we can choose the input dimension $k$ to be larger than $n_\mu + 1$.


Following a recent line of works on nonlinear model reduction with neural networks \cite{lee2020model, choi, tencer2021tailored, romor2022non}, we use autoencoders for learning the map $\hat{\vg}$. Departing from this line of works though, and inspired by the variational autoencoders \cite{kingma2013auto}, we consider \textit{stochastic noisy encoders} $\hat{\ve} : \R^{n_s} \to \R^{k}$. Specifically, given an input vector ${\vy_s} \in \R^{n_s}$ the noisy encoder outputs two vectors corresponding to the mean and standard deviation of a gaussian latent vector

\[
	\hat{\ve}({\vy_s}) = \bm{\eta}({\vy_s}) + \bm{\sigma}({\vy_s}) \odot \bm{\xi} \quad \bm{\xi} \sim \mathcal{N}(0, \mI_k)
\]
where $\odot$ denotes the entrywise multiplication.
The decoder is then trained to reconstruct the original vector $\vy_s$ from the stochastic latent vectors outputs of the encoder
\[
	\hat{\vg}( \hat{\ve}(\vy_s) ) \approx \vy_s.
\]

To train the encoder and decoder networks we use stochastic gradient descent so that for each batch $\{ \vy^{j} \}_{j=1}^b \subset \mathcal{Y}_{\text{train}}$ we consider the loss 
\begin{equation}\label{eq:loss_ae}
	L(\Theta_e, \Theta_g) := \frac{1}{2} \sum_{j = 1}^b {\| \hat{\vg}(  \hat{\ve}(\mP_s \vy^j)) - \mP_s \vy^j \|_2^2} \big/ \sum_{j = 1}^b {\| \mP_s \vy^j \|_2^2}
\end{equation}
where $\Theta_e$ and $\Theta_g$ are set of learnable parameters of the encoder and decoder networks respectively, and $\mP_s \in \R^{n_s \times N}$ restricts the vectors $\{\vy^j\}_j$ on the stencil mesh. 

Using a noisy encoder to train the decoder network can be interpreted as a regularization method that encourages the decoder to be stable to perturbations. We empirically observe that at the end of the training, the encoder tends to output vectors $\sigma_y$ with negligible norms, making it effectively deterministic during inference.  Nonetheless, the noisy autoencoders we find have low training and test error with respect to autoencoders trained without noisy encoders. Furthermore, we observe that training is more stable with respect to random initialization, that is networks trained from different random initialization have all similar performances, contrary to the standard autoencoders where performances may vary significantly from one run to the other. 
See the numerical experiments for more details.

   	\subsection{Full state reconstruction via Gappy-POD}\label{subsec:gappyPOD}

The trained hyper-reduced decoder $\hat{\vg}$ found in the previous step can now be used in the manifold LSPG with collocation \eqref{eq:hyperODeltaE_hat} to estimate the restriction of the FOM states $\vy^n$ on the stencil mesh. 

Let $\hat{\vy}^n = \hat{\vg}(\vz^n)$ be an estimate of $\mP_s \vy^n$ and $\bm{\Phi}_r = \bm{\Phi}[:, :r]$ a restriction of the POD matrix $\mPhi$ to the first $r$ principal modes, we propose to finally obtain an approximation $\vy^n$ by the Gappy-POD method \cite{everson1995karhunen}
\begin{equation}\label{eq:GappyApprox}
	\vy^n \approx \bm{\Psi} \hat{\vy}^n \quad\text{with}\quad \bm{\Psi} := \bm{\Phi}_r  \big( \mP_s \bm{\Phi}_r \big)^\dagger
\end{equation}
where for a matrix $\mA \in \R^{m \times n}$ we denote by $\mA^\dagger \in \R^{n \times m}$ its Moore-Penrose inverse. 

Notice that the application of the matrix $\bm{\Psi}$ to the output of the decoder $\hat{\vg}$ can be interpreted as extending $\hat{g}$ by adding a linear layer with weight $\bm{\Psi}$. Taking this idea a step further, we can add nonlinearities to \eqref{eq:GappyApprox} in order to encode prior knowledge on the solutions $\{ \vy^n \}$ of the FOM \eqref{eq:FOM}. For example, knowing that the solutions of the FOM are positive, and following \cite{romor2022non}, we can enforce the positivity of the approximations by adding a $\relu$ nonlinearity to the outputs of the Gappy-POD method:
 \begin{equation}\label{eq:GappyApprox_wRelu}
	\vy^n \approx \relu(\bm{\Psi} \hat{\vy}^n)
\end{equation}
with $\bm{\Psi}$ as in \eqref{eq:GappyApprox}. In the next sections we refer to the latter model as \textit{hyper-decoder + Gappy-POD + $\relu$}.
\section{Numerical Results: 2d Burgers equation}\label{sec:num_res}

We demonstrate our approach on the 2D Viscous Burgers' equation 
\begin{align*}
 \frac{\partial u}{\partial t} + \frac{1}{2} \frac{\partial u^2}{\partial x} + \frac{1}{2} \frac{\partial u v}{\partial y}  &= \frac{1}{Re} \left( \frac{\partial^2 u}{\partial x^2} + \frac{\partial^2 u}{\partial y^2} \right) \\
   \frac{\partial v}{\partial t} + \frac{1}{2} \frac{\partial u v}{\partial x} + \frac{1}{2} \frac{\partial v^2}{\partial y}  &= \frac{1}{Re}\left( \frac{\partial^2 v}{\partial x^2} + \frac{\partial^2 v}{\partial y^2} \right)\\
   (x,y) &\in \Omega = [0,1] \times [0,1],\\
   t &\in [0,2]
 \end{align*}
 with periodic boundary conditions and parameterized initial conditions
 \begin{align*}
 	u(x,y,0) = v(x,y,0) = \mu \sin(2 \pi x) \cdot \sin(2 \pi y) \chi_{[0, 0.5]^2}.
 \end{align*}
 Here $Re$ is the Reynolds number, $u, v$ denote the velocities in the $x$ and $y$ direction, $\mu$ is the scaling parameter for the initial conditions, and for a set $\mathcal{S} \subset \Omega$ we denote by $\chi_{\mathcal{S}}$ is the indicator function over the set $\mathcal{S}$.
 
 \subsection{Full-Order Model}

We use the open-source Python libary Pressio-Demoapps \footnote{https://pressio.github.io/pressio-demoapps/index.html} to solve the previous 2D Burgers' equation. The equation is discretized in space using the finite volume method on a structured orthogonal $60 \times 60$ grid with a stencil size of 3. 

We collect the FOM solutions for different values of parameters $\mu \in \mathcal{D}_{\text{train}} \cup \mathcal{D}_{\text{validation}} \cup \mathcal{D}_{\text{test}}$ using Runge-Kutta4 for time integrator with $N_t = 2000$ time steps of size $\Delta t  = 10^{-3}$. In the first set of experiments we take 
\begin{equation}\label{eq:Dtrain}
\mathcal{D}_{\text{train}} := \{0.16 , 0.208, 0.256, 0.304, 0.352, 0.448, 0.496, 0.544, 0.592, 0.64\}
\end{equation}
to construct the training set. We also consider a smaller training set where $\mu$ is taken in 
\begin{equation}\label{eq:Dtrainp}
	\mathcal{D}_{\text{train}}' := \{0.16, 0.28, 0.52, 0.64\}.
\end{equation}
We use the snapshots corresponding to $\mu \in \mathcal{D}_{\text{validation}} = \{0.4\}$ for validation. To save storage we collect the training snapshots every 4 steps, leading to a total number of 5003 training datapoints for $\mathcal{D}_{\text{train}}$ and 2001 for $\mathcal{D}_{\text{train}}'$. The test set is instead constructed by taking 34 equispaced test parameters from the range $[0.06, 0.72]$. Note that some of these parameters are outside the training interval $[0.16, 0.64]$ and will be used to study the extrapolation performance of the proposed method.

   	\subsection{Sample and stencil meshes}

We use the methods described in Section \ref{subsec:stencil_mesh} to construct the sample and stencil meshes. Notice that because of the symmetries of the problem and of the initial conditions, the values of the velocities $u$ and $v$ are always identical, i.e. $u(x,y,t) = v(x,y,t)$ for all $(x,y)\in \Omega , t \in [0,T]$. For this reason, we construct a datamatrix $\mU$ by selecting only the rows corresponding to the variable $u$ in $\mY_{\text{train}}$. We then let $\mPhi^{(u)}$ be the orthogonal (POD) matrix obtained via the Singular Value Decomposition (SVD) of $\mU$. 

\begin{figure}
      \centering
      \subcaptionbox{$n_c = 50$ ($n_s = 161$)\label{fig:nc50}}
        {\includegraphics[scale=0.5]{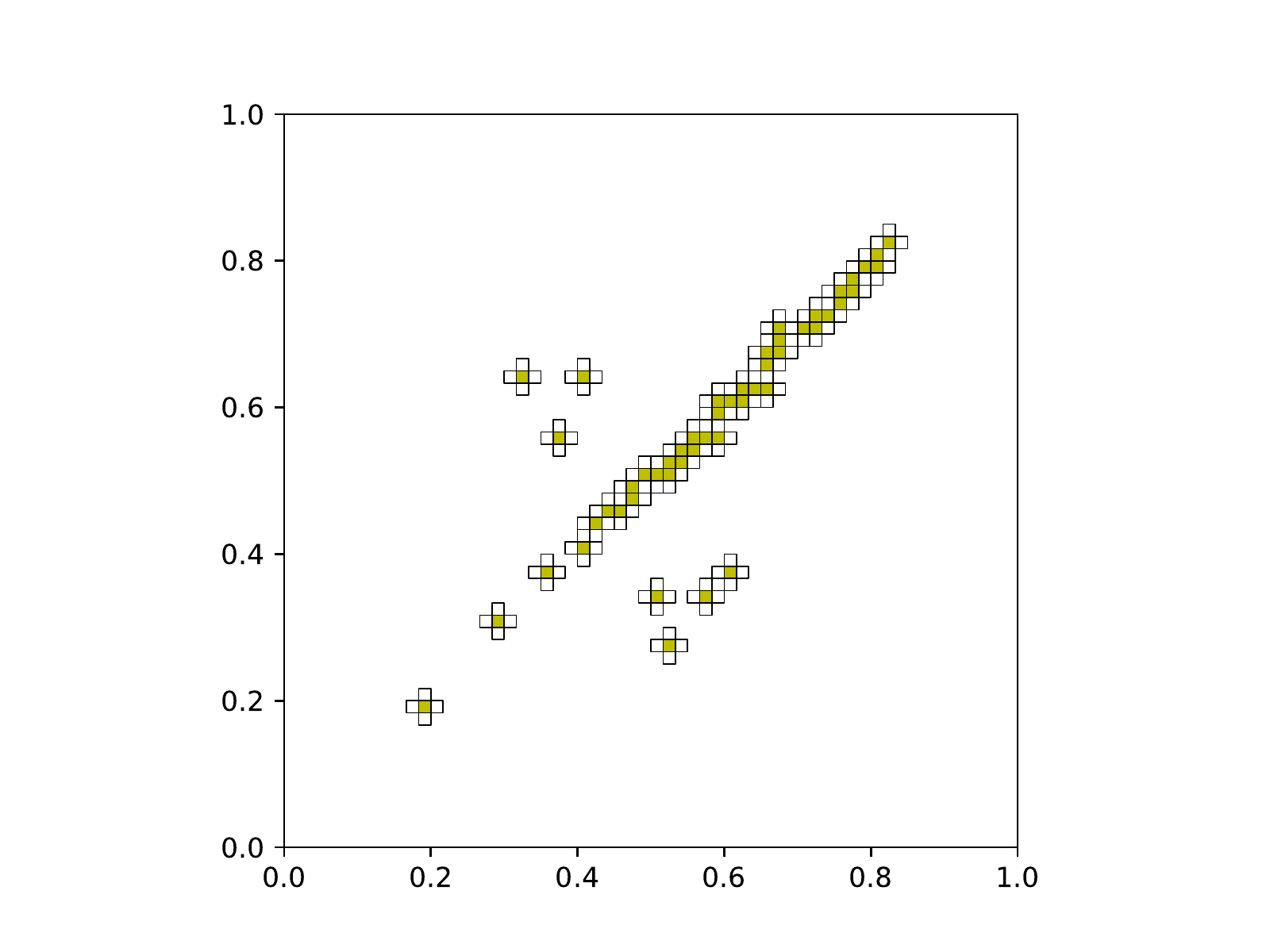}}
        \hskip -6ex
       \subcaptionbox{$n_c = 100$ ($n_s = 296$)\label{fig:nc50}}
        {\includegraphics[scale=0.5]{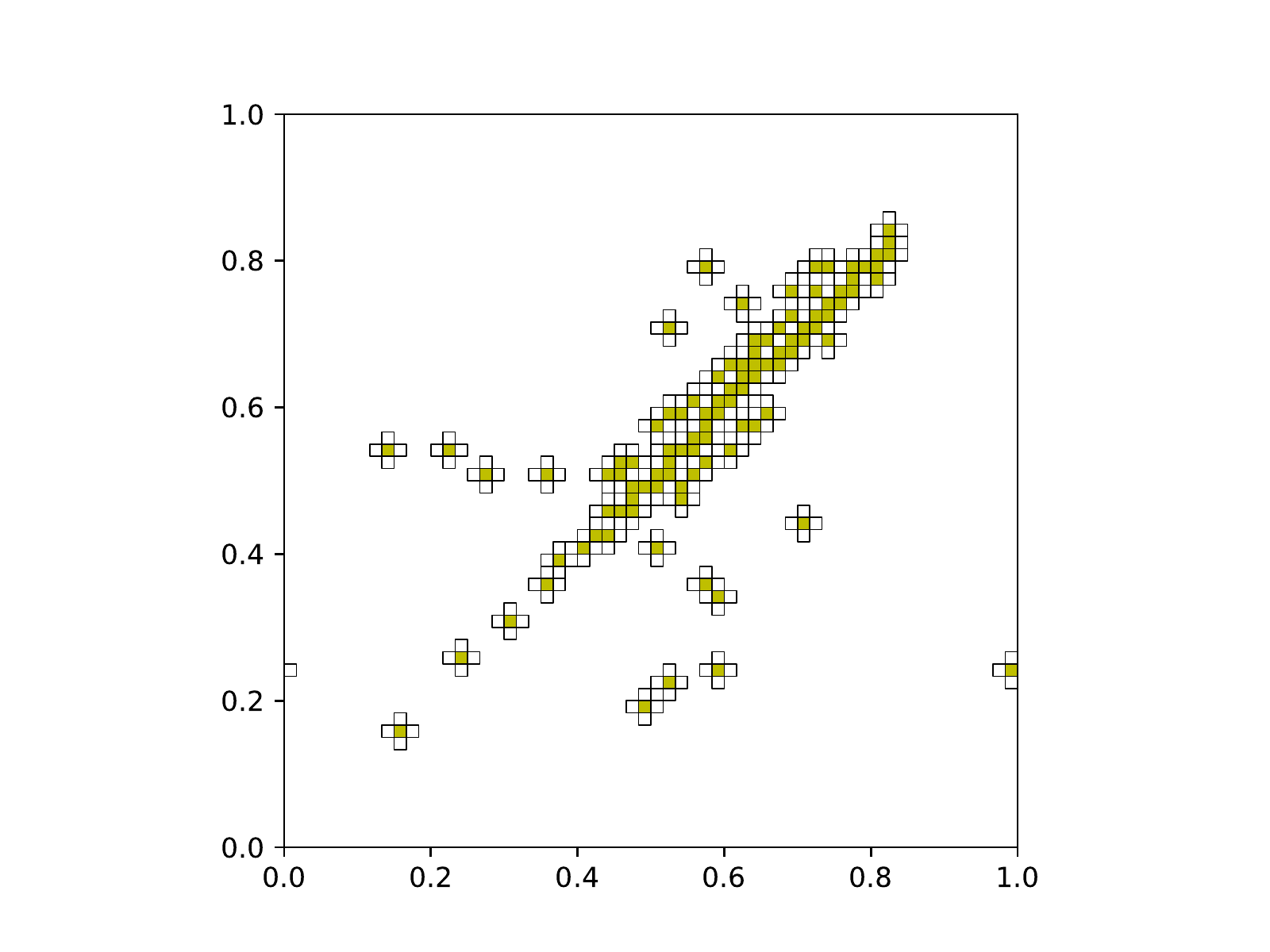}}
      \hskip -6ex
       \subcaptionbox{$n_c = 150$ ($n_s = 440$)\label{fig:nc50}}
        {\includegraphics[scale=0.5]{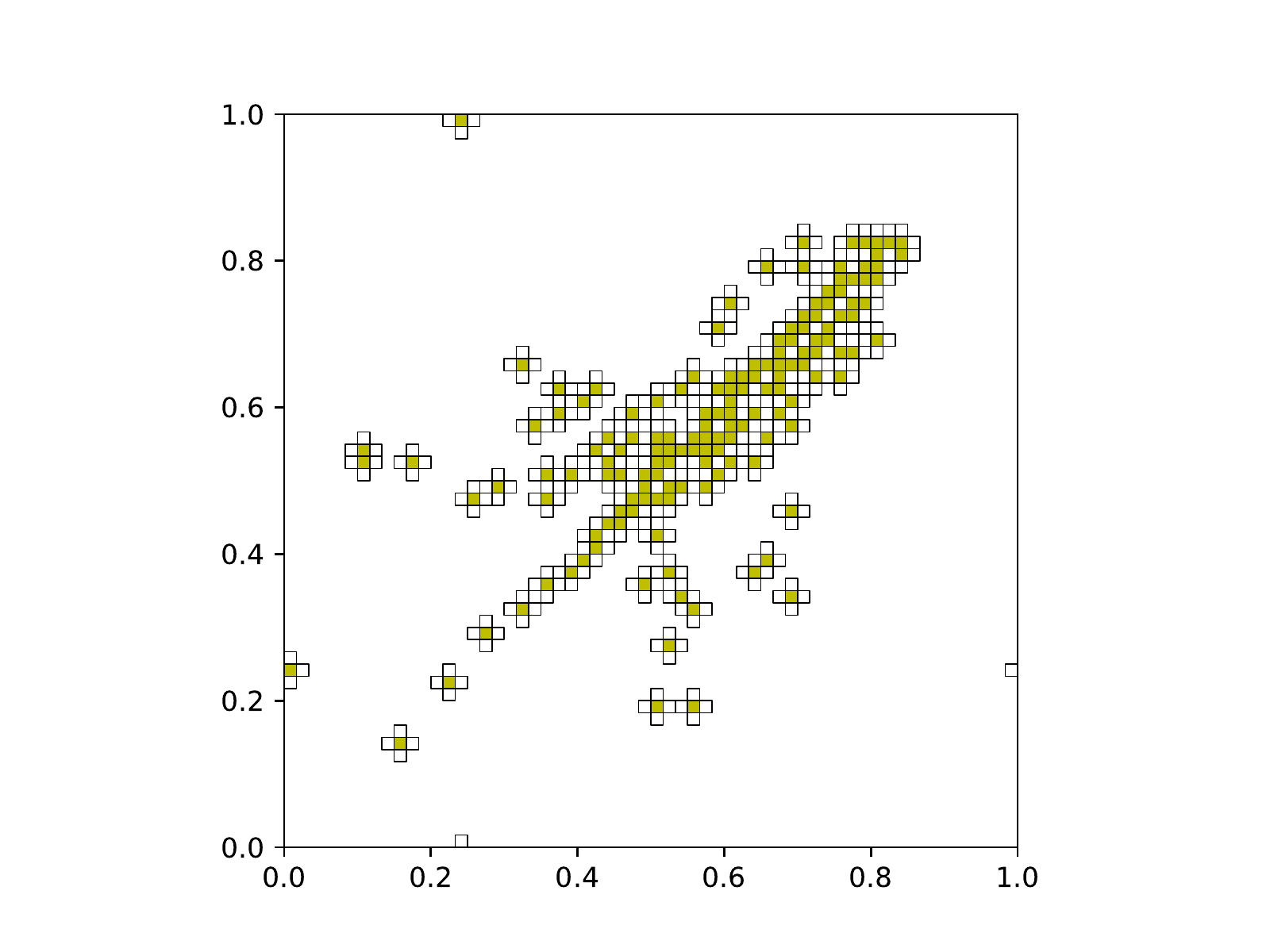}}
      \caption{Sample and stencil meshes for a varying number of $n_c$ number of sample mesh points obtained via Algorithm \ref{algo:QRSampling}. The cells shown are those in the stencil mesh, the ones in yellow are those in the sample mesh. With $n_s$ we denote the number of cells in the stencil mesh.}\label{fig:meshes}
\end{figure}

We apply Algorithm \ref{algo:QRSampling} with $\mPhi = \mPhi^{(u)}$  and sample mesh sizes $n_c \in \{ 50, 100, 150 \}$. Notice that the algorithm requires only the number of cells $n_c$ to be included in the sample mesh, the number of cells $n_s$ in the stencil mesh depends on the sample mesh and the discretization used. In this case, since we used stencil size of 3 for the space discretization, the inclusion of an index of a cell in the sample mesh implies the inclusion of also the indexes of the four adjacent cells in the stencil mesh. In Figure \ref{fig:meshes} we plot the stencil mesh where the yellow cells denote the cells in the sample mesh.

   	\subsection{Nonlinear manifold learning}

Since we have two variables $(u,v)$ the dimension of the FOM state $\vy$ is twice the one of the grid size. For the same reason, the dimension of the output of the \textit{hyper-decoder} and the input of the noisy-encoder is twice the dimension of the stencil size ($2 n_s$), i.e. $\hat{\vg}: \R^{k} \to \R^{2 n_s}$ and~$\hat{\ve}: \R^{2 n_s} \to \R^{k}$.  

Notice that even for the case when $n_c = 150$, the corresponding size of the input/output of the networks ($2 n_s$) is only a fraction of the size of the FOM state. This will allow us to train a single neural network for the two variables, saving computational resources. This is in contrast to recent works, as for example \cite{choi}, which define autoencoders on the full mesh and, to reduce the memory costs, train separate networks for each variable. Finally, training a single neural network for multiple variables allows for reducing the total size of the ROM state (latent vectors) forcing the decoder to learn a meaningful and shared representation of the data.

\medskip 

For all the experiments, we consider shallow feed-forward encoders and decoders with fixed architectures. Specifically, for and $\vy_s \in \R^{2 n_s}$ the forward pass in the noisy-encoder is given by
\[
\begin{aligned}
	\vh &= \relu(\mW_1^{(e)} \vy_s + \vb_1), \\
	\bm{\eta} &= \mW_{2,\eta}^{(e)} \vh + \vb_{2,\eta} \\
	\bm{\sigma} &= \exp((\mW_{2,\sigma}^{(e)},\vh + \vb_{2,\sigma})/(2*\ell)), \\
	\hat{\ve}(\vy_s) 
	&= \bm{\eta} +  \bm{\sigma} \cdot \bm{\xi} \quad\text{with}\quad \bm{\xi} \sim \mathcal{N}(0, \mI_k)
\end{aligned}	
\]
where $\exp$ and $\relu(z) = \max(0, z)$ are applied entrywise. The scalar $\ell$ is a hyperparameter that regulates the size of the noise. Tuning it on our dataset we find that $\ell = 1$ gives the best results and we fix it for all the next experiments. The parameters of the encoders are $\mW_1^{(e)} \in \R^{100 \times 2 n_s}, \vb_1 \in \R^{100}, \mW_{2,\eta}^{(e)} \in \R^{5 \times 100}, \vb_{2,\eta} \in \R^{5}$ and $\mW_{2,\sigma}^{(e)} \in \R^{100 \times 5}, \vb_{2,\sigma} \in \R^{5}$.

\medskip

The hyper-decoder is a two-layers $\leakyRelu$ network. For a latent vector $\vz \in \R^k$ we have
\[
	\hat{\vg}(\vz) = \mW_2^{(d)} \leakyRelu ( \mW_1^{(d)} \vz),
\]
where $\mW_1^{(d)} \in \R^{700 \times k}$ and $\mW_2^{(d)} \in \R^{2 n_s \times 700}$ are the parameters of the decoder, and $\leakyRelu(z) = \max(0, z) + 0.01 \min(0, z)$ is applied entrywise. 

When solving the time discrete ROM \eqref{eq:hyperODeltaE_hat}, we need to compute the Jacobian of the hyper-decoder. We compute this Jacobian via its closed form, which is faster than using automatic differentiation or approximating it with finite differences, and is given by
\[
	\mJ_{g} (\vz) = \mW_2^{(d)} \big[ \text{diag}(\mW_1^{(d)} \vx > 0) + 0.01 \text{diag}(\mW_1^{(d)} \vz < 0) \big] \mW_1^{(d)}
\]

\medskip

We train the encoder and decoder networks using PyTorch Lightning \cite{Falcon_PyTorch_Lightning_2019}, an open-source library of PyTorch.
 We use ADAM \cite{kingma2014adam} (a variant of stochastic gradient descent)  for 2000 epochs with a batch size of 120. The learning rate is initialized at $0.005$ and reduced by a factor of 0.8 if the training loss does not improve for 20 epochs until a minimum learning rate of $1e$-7. The network weights are initialized via the standard PyTorch initialization.

   	\subsection{Nonlinear manifold LSPG with collocation	} 

The trained hyperdecoder can now be used in the nonlinear manifold LSPG formulation \eqref{eq:hyperODeltaE_hat}. The residual is defined by choosing for time discretization the implicit methods BDF1 and BDF2. 

 The manifold LSPG is solved using the python interface of Pressio \cite{rizzi2020pressio}, an open-source library for model-reduction that enables large-scale projection-based reduced order models for nonlinear dynamical systems. The minimization problem \eqref{eq:hyperODeltaE_hat} is solved with the Gauss-Newton method, where Pressio-demoapps is used to provide Pressio with the residual $\vr$ and its Jacobian $\mJ_r$ evaluated on the stencil mesh.
 
Solving the ROM \eqref{eq:hyperODeltaE_hat} for a given parameter $\mu$, we obtain the ROM states $\{ \vz^n_\mu \}_{n}$, and estimate the FOM states $\{ \vy^n_\mu \}_{n}$ by $\tilde{\vy}^n_\mu := \relu(\bm{\Psi} \hat{\vg}(\vz^n_\mu))$ where $\bm{\Psi}$ is the Gappy matrix defined in \eqref{eq:GappyApprox} (hyper-decoder + Gappy-POD + $\relu$). The number $r$ of POD modes used to define $\bm{\Phi}_r$ is chosen by finding the number of modes $r$ that perform the best on the training and validation set. 
 
\begin{figure}[h]
    \centering 
\begin{subfigure}{0.3\textwidth}
  \includegraphics[width=\linewidth]{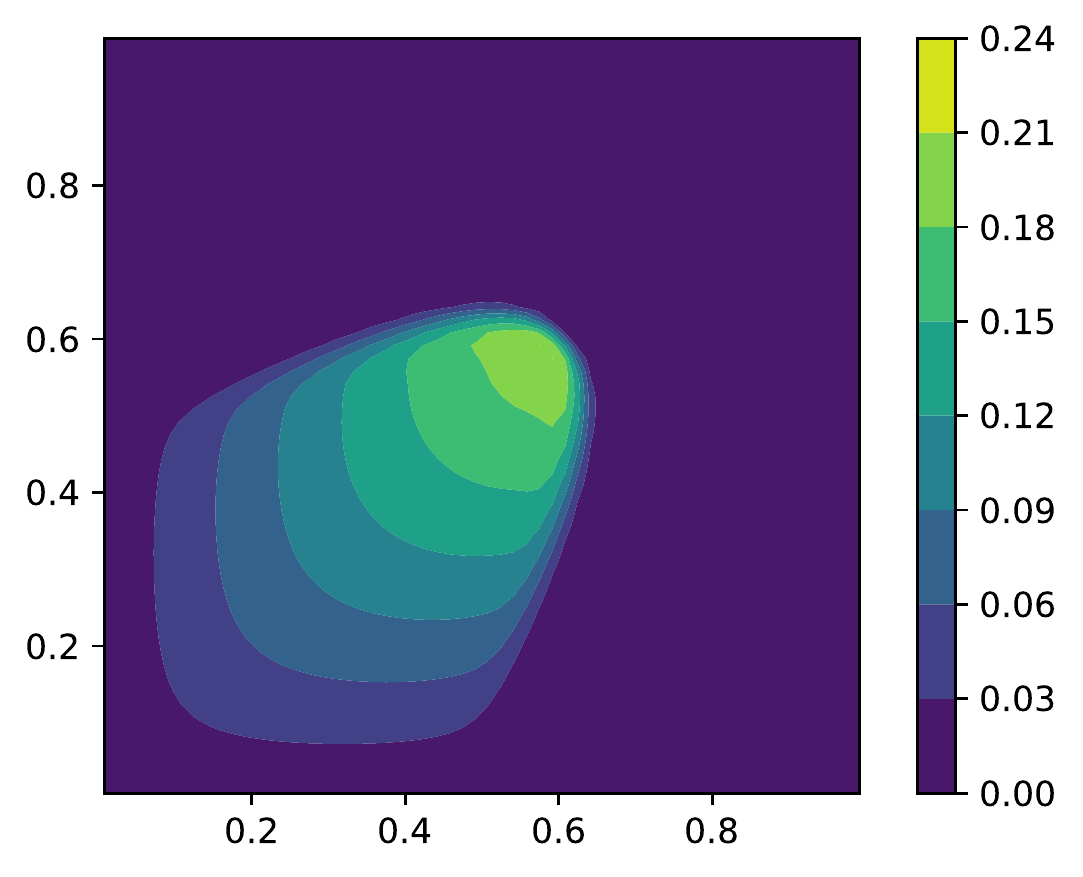}
  \caption{Target $u$ ($\mu = 0.26$)}
\end{subfigure}\hfil 
\begin{subfigure}{0.3\textwidth}
  \includegraphics[width=\linewidth]{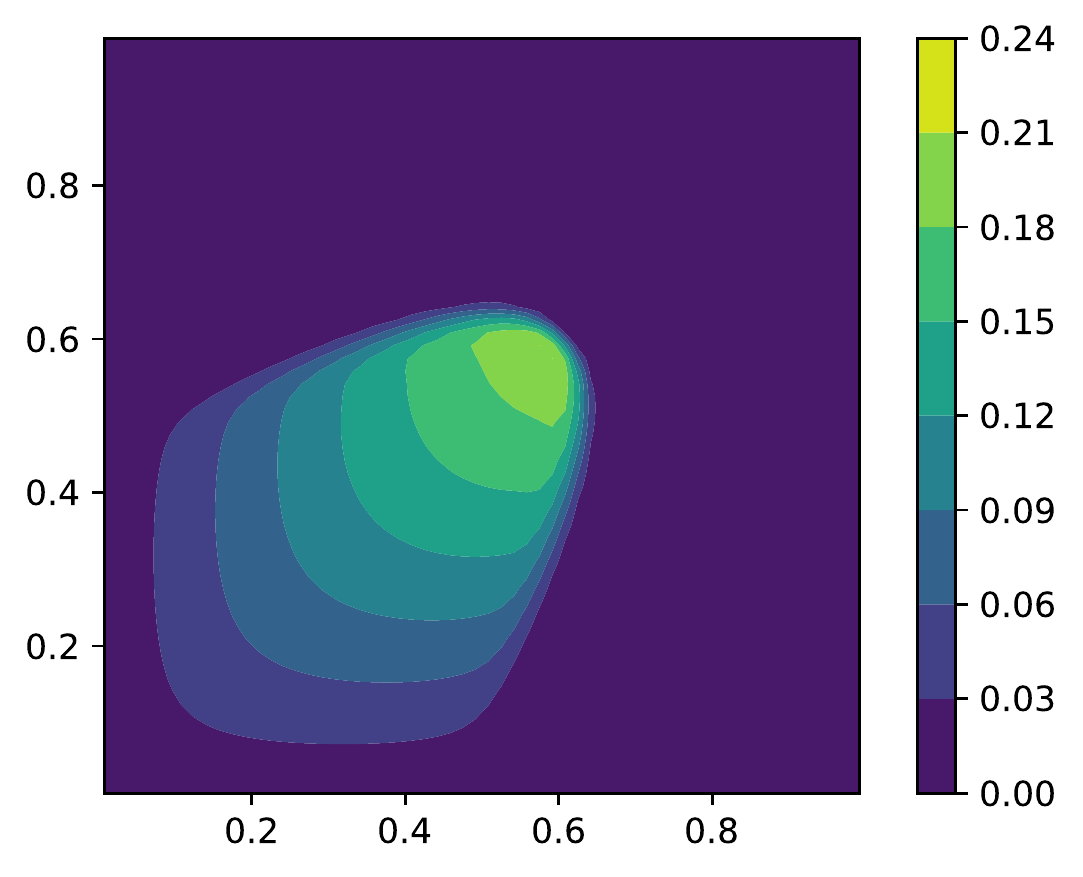}
  \caption{Predictions $u$ ($\mu = 0.26$)}
\end{subfigure}\hfil 
\begin{subfigure}{0.3\textwidth}
  \includegraphics[width=\linewidth]{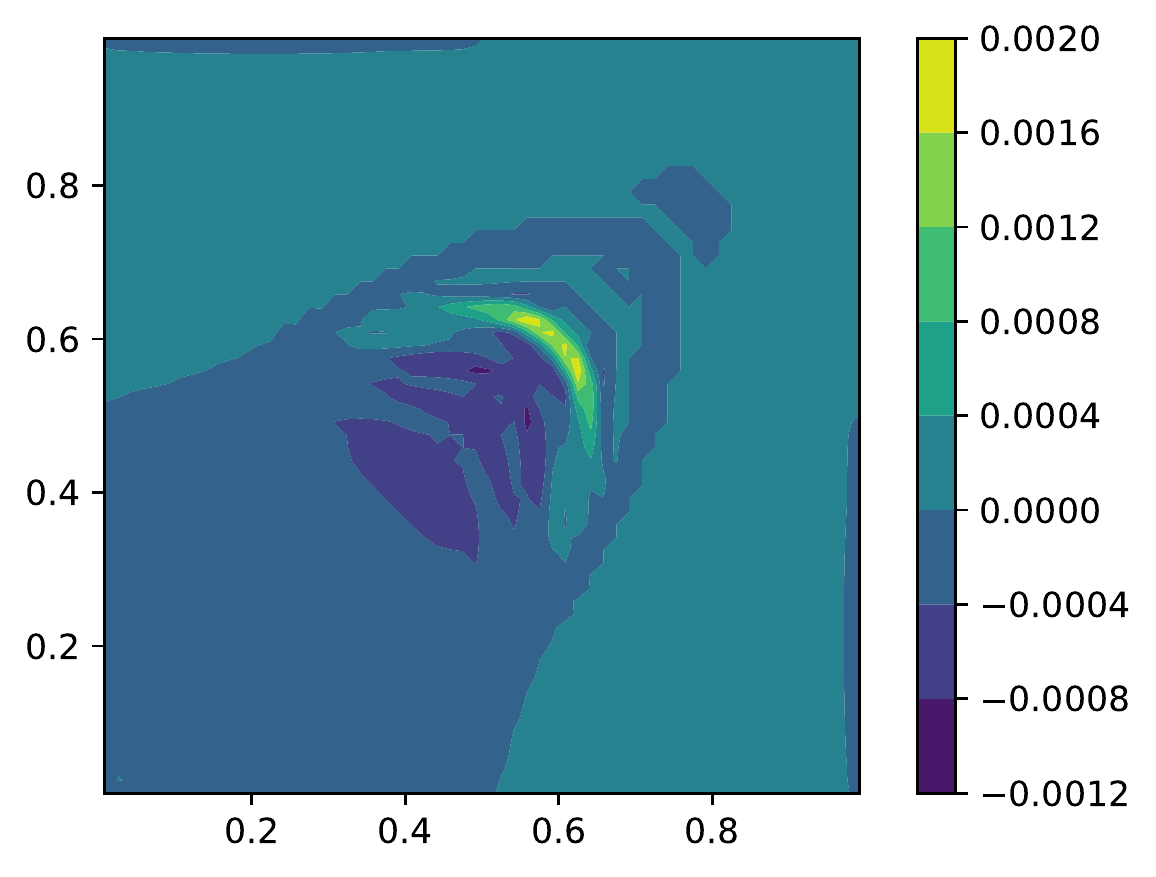}
  \caption{Error (target - predictions)}
\end{subfigure}
\medskip
\begin{subfigure}{0.3\textwidth}
  \includegraphics[width=\linewidth]{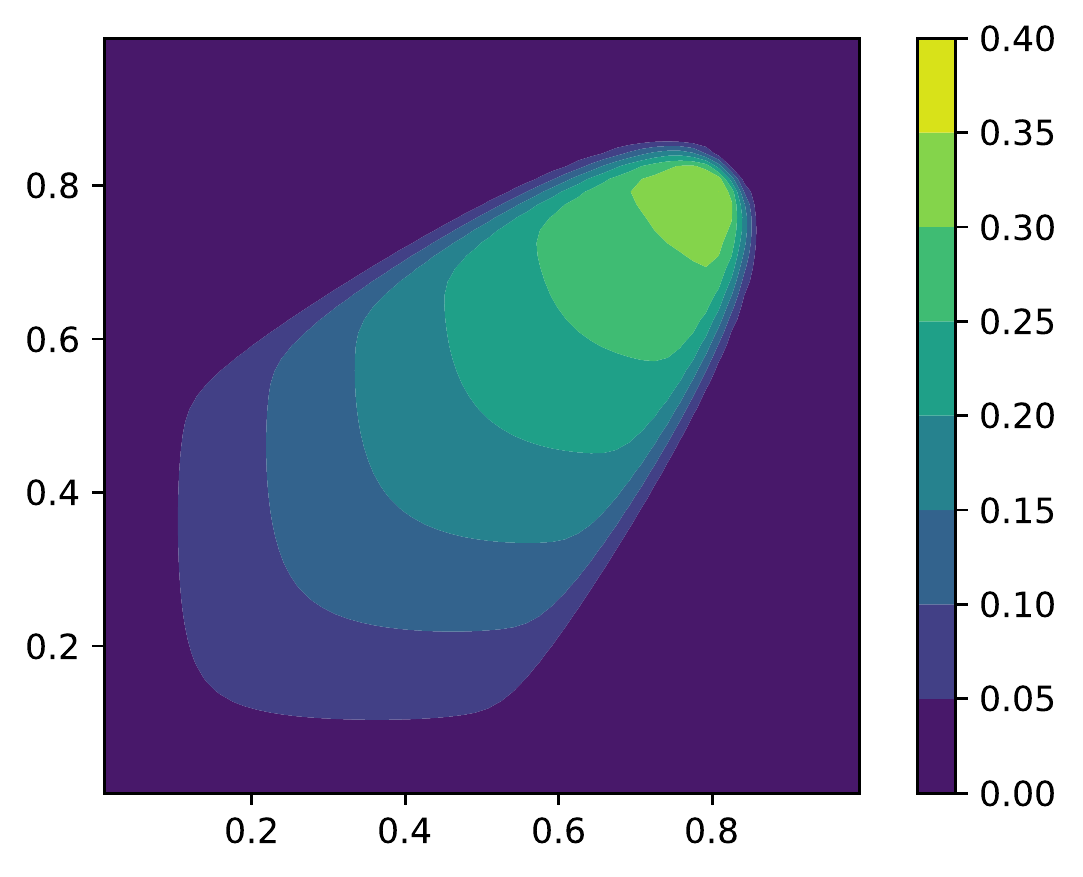}
  \caption{Target $u$ ($\mu = 0.62$)}
\end{subfigure}\hfil 
\begin{subfigure}{0.3\textwidth}
  \includegraphics[width=\linewidth]{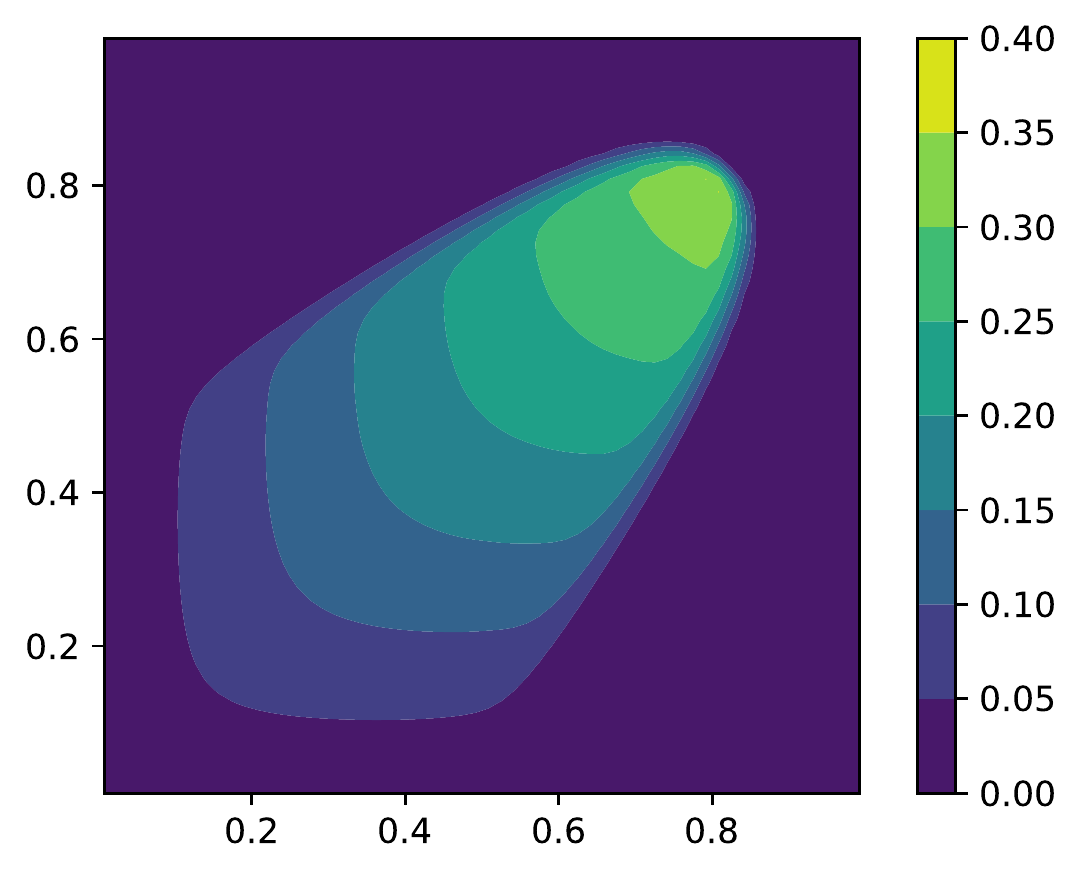}
  \caption{Predictions $u$ ($\mu = 0.62$)}
\end{subfigure}\hfil 
\begin{subfigure}{0.3\textwidth}
  \includegraphics[width=\linewidth]{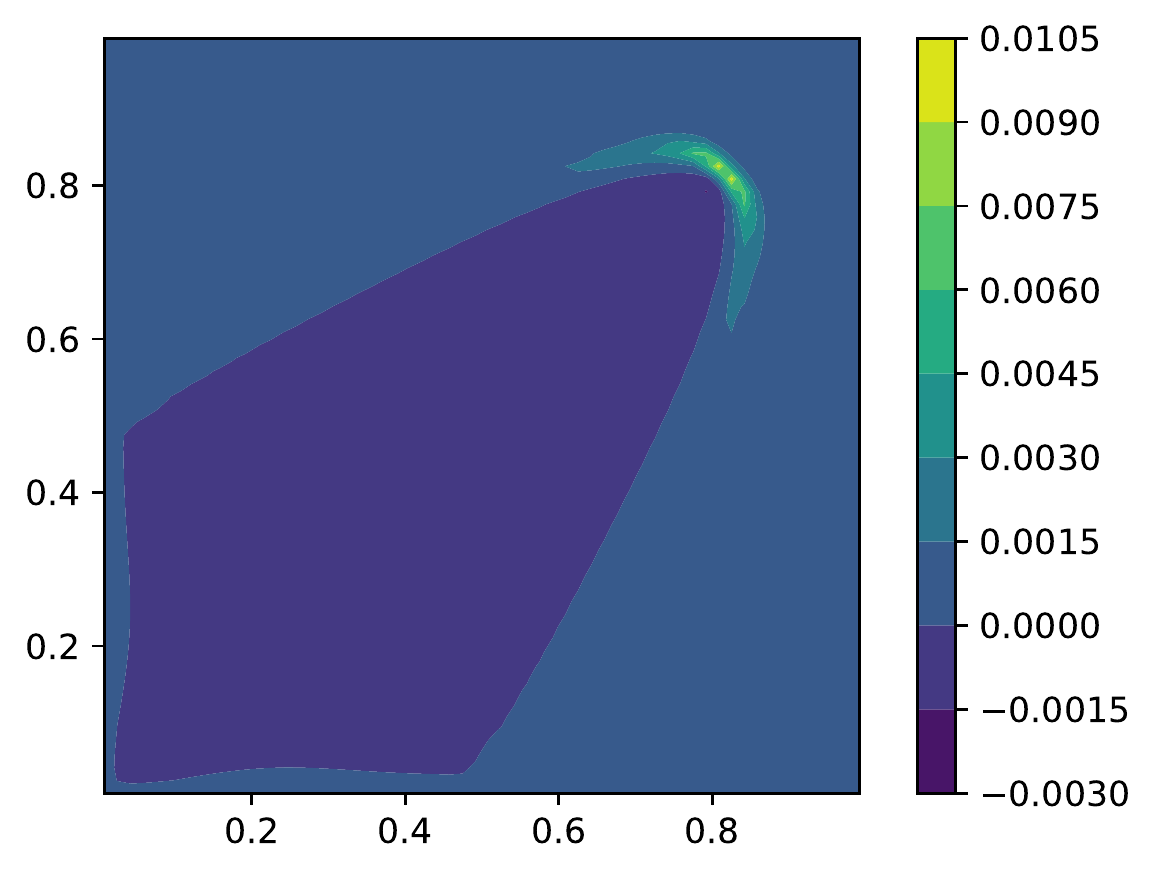}
  \caption{Error (target - predictions)}
\end{subfigure}
\caption{Comparison of the target final states with the predictions via Gappy-POD + hyperdecoder, and corresponding errors (target - predictions) fore two target parameters and sample-mesh size $n_c = 50$.}
\label{fig:qualitative}
\end{figure}

 In Figure \ref{fig:qualitative} we compare the final state predictions with the target solutions for the variable $u$, for two different parameters in the test set ($\mu = 0.26$ and $\mu = 0.62$) and with a sample mesh size of $n_c = 50$. The errors are relatively low across the whole domain, increasing moderately at the wavefront of the solutions.
 
 In the next sections, we evaluate the performance on test cases with $\mu \in \mathcal{D}_{\text{test}}$, using the following two metrics: 
 \[
 	\text{Maximum relative error} := \max_{n \in [N_t]} \Bigg( \frac{\| {\vy}^n_\mu - \tilde{\vy}^n_\mu\|_2}{\|{\vy}^n_\mu\|_2} \Bigg)
 \]	
 and 
  \[
 	\text{${L}^2$ relative error} :=  \frac{\sum_{n \in [N_t]} \| {\vy}^n_\mu - \tilde{\vy}^n_\mu\|_2}{\sum_{n \in [N_t]} \|{\vy}^n_\mu\|_2},
 \]	
 where for a natural number $n$ we let $[n] := \{1,2, \dots, n\}$.

 \subsection{Effects of the noisy-training}
 
In this section, we consider the effects of training via the noisy autoencoder described in Section \ref{subsec:AE}. 
We train 14 models using noisy autoencoders from different random initialization and 14 models using standard encoders. 
The models have essentially the same encoder-decoder architecture, 
the only difference being that the models in the first group perturb the output of the encoder 
with Gaussian variables with learnable variance. 

\begin{figure}[h]
      \centering
        {\includegraphics[scale=0.7]{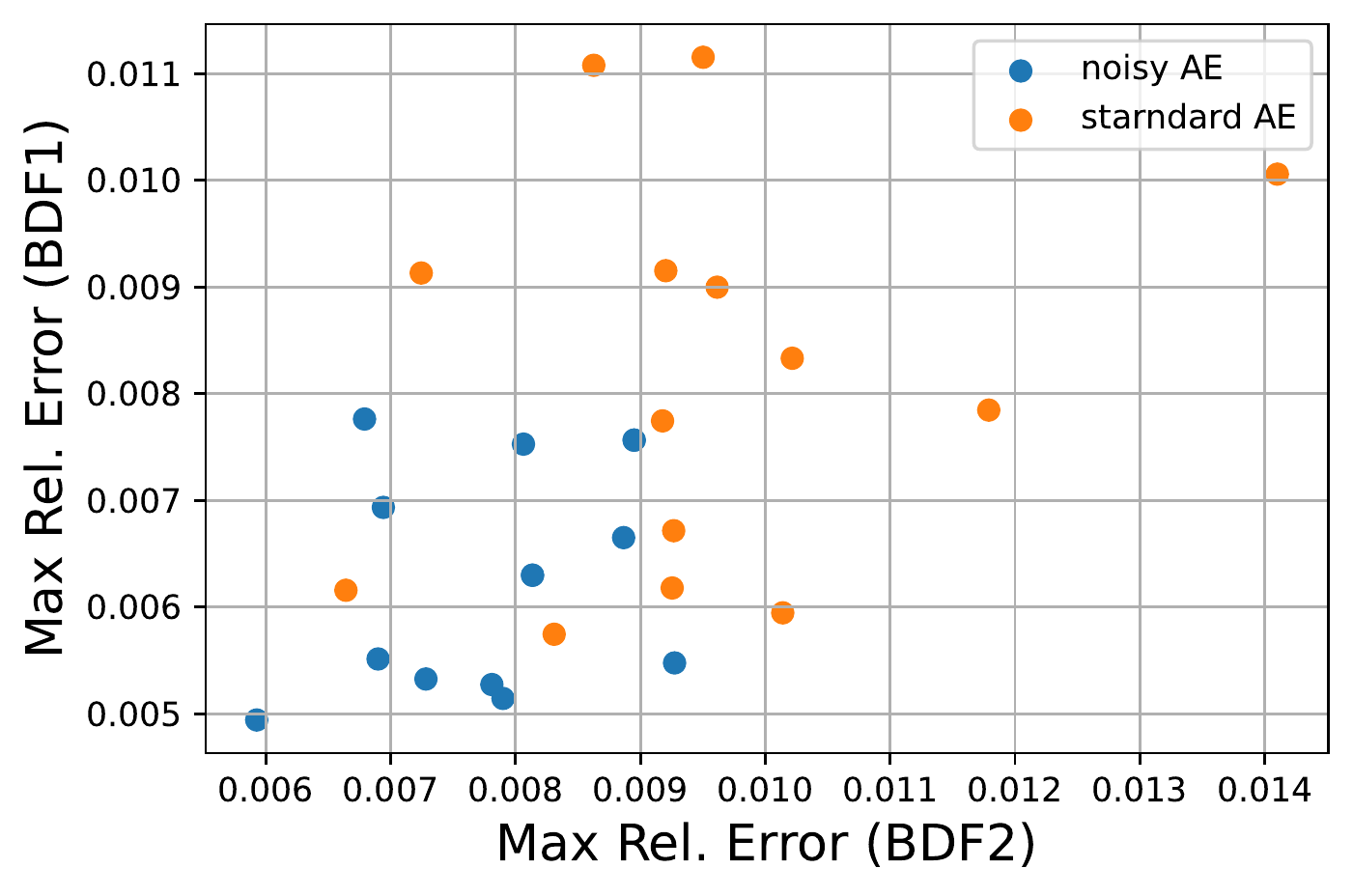}}
            \caption{Maximum relative errors for models trained with noisy and standard autoencoders.}\label{fig:scatter}
\end{figure}

We consider a sample mesh of $n_c = 50$ and use the trained models to solve 
manifold LSPG with collocation for the validation parameter $\mu = 0.4$. 
In Figure \ref{fig:scatter} we plot the maximum relative errors using BDF1 and BDF2 
time integration. Notice that the models trained with noisy autoencoders cluster in the 
lower left side of the plane where the errors with BDF1 and BDF2 are both low.

During the training of the noisy autoencoder, we observe that the standard deviation 
$\sigma(\vy)$ of the noisy encoder decreases progressively until it becomes negligible 
at the end of training. This suggests that the noisy autoencoder can be viewed as a form 
of adaptive noisy training, where the level of noise decreases to zero over time. 
The use of noise injection as a regularization method has been studied extensively in prior 
work (e.g., \cite{bishop1995training, poole2014analyzing, dhifallah2021inherent}).
	
Based on these observations, we use noisy autoencoders with the noise term set to zero in the remaining experiments. 
	
 \subsubsection{Effects of the sample mesh size $n_c$}
 
\begin{figure}[h]
      \centering
        {\includegraphics[scale=0.5]{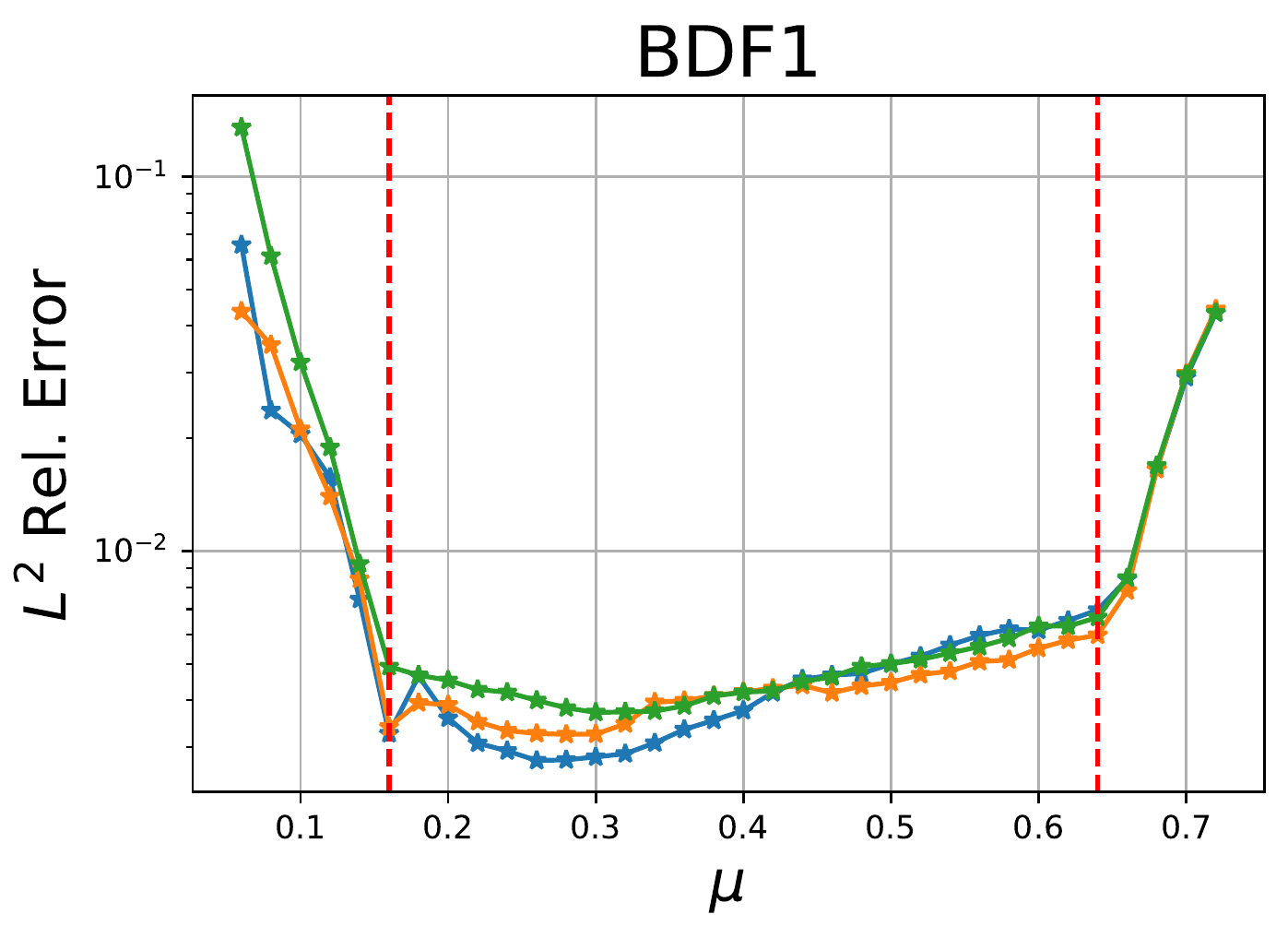}}
        {\includegraphics[scale=0.5]{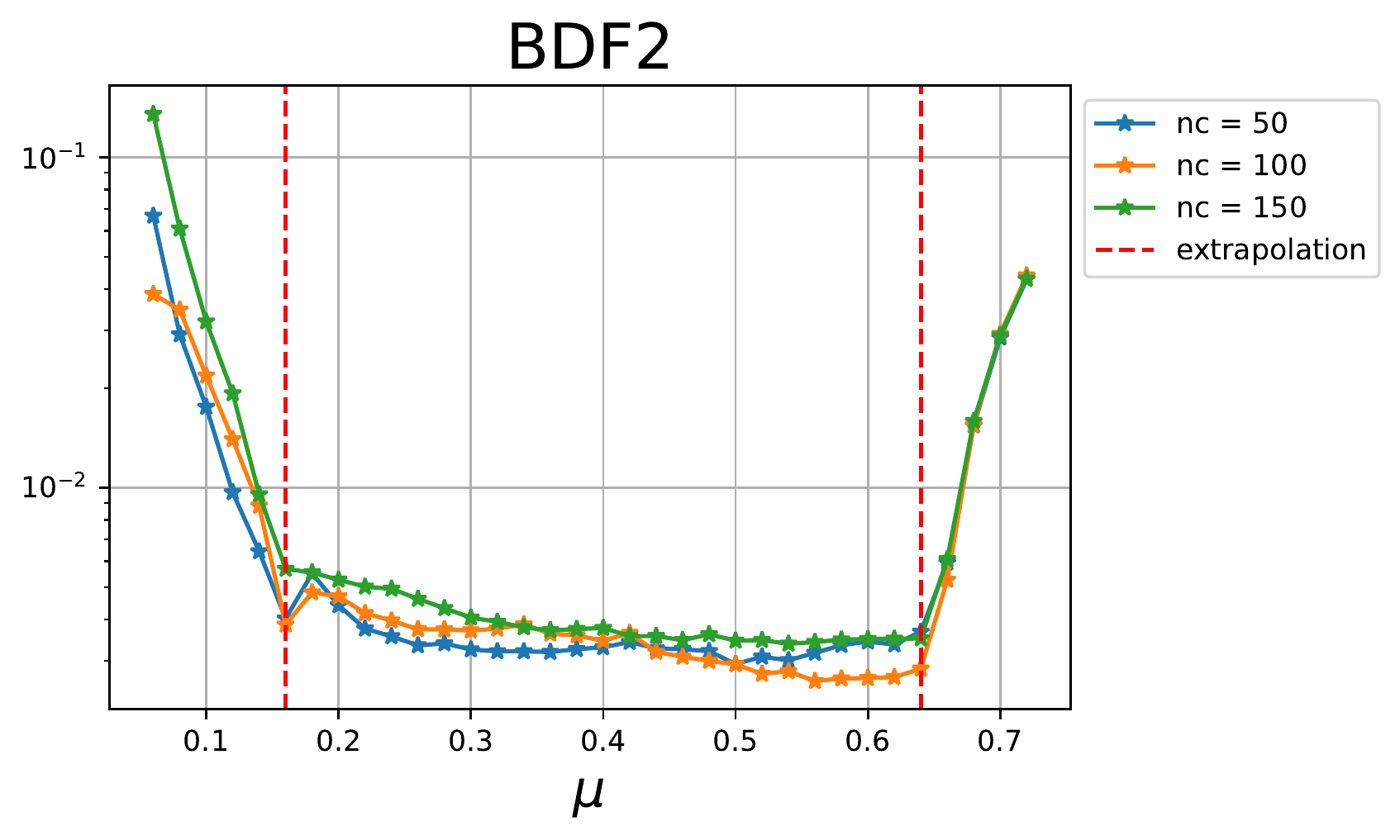}}
        {\includegraphics[scale=0.5]{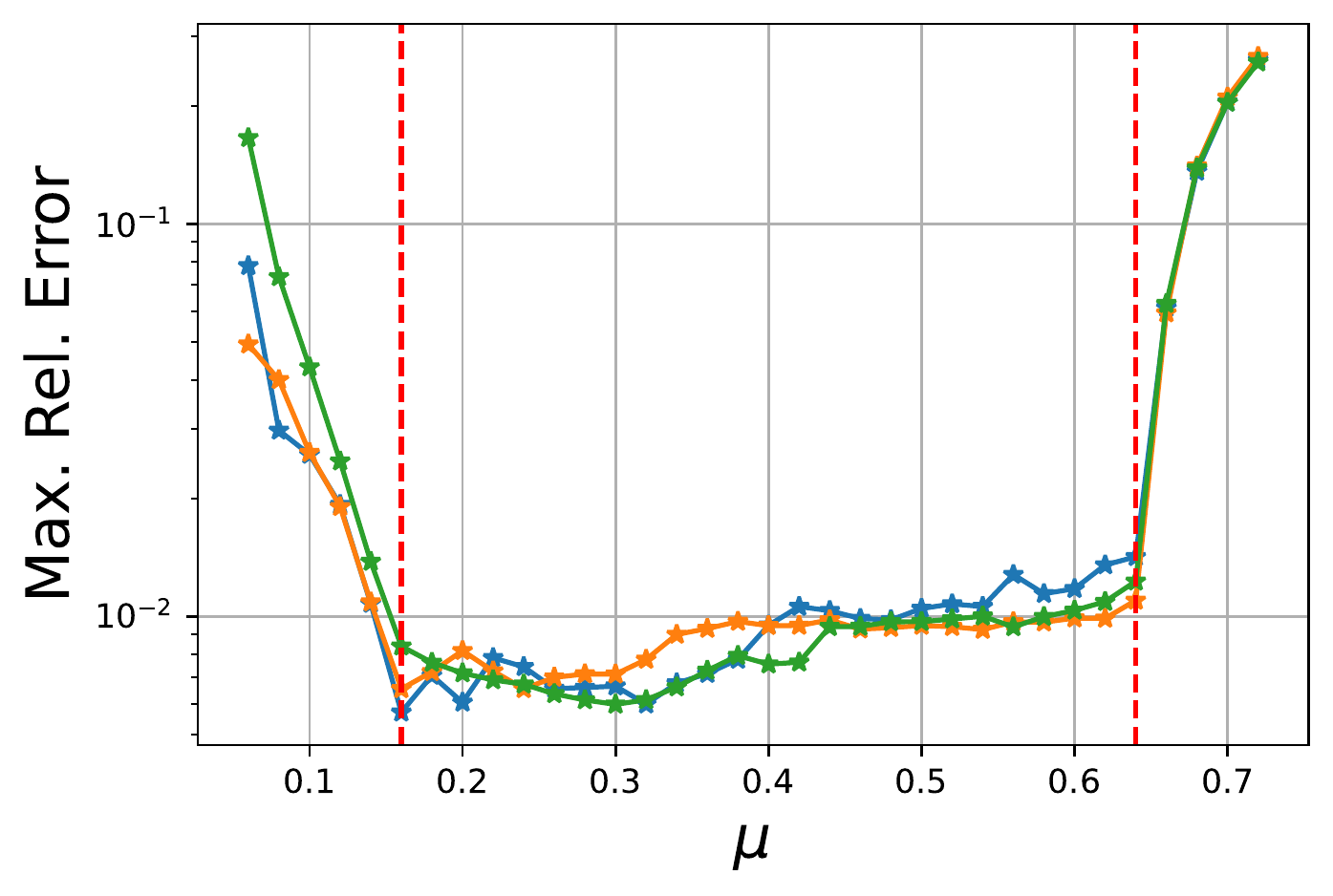}}
        {\includegraphics[scale=0.5]{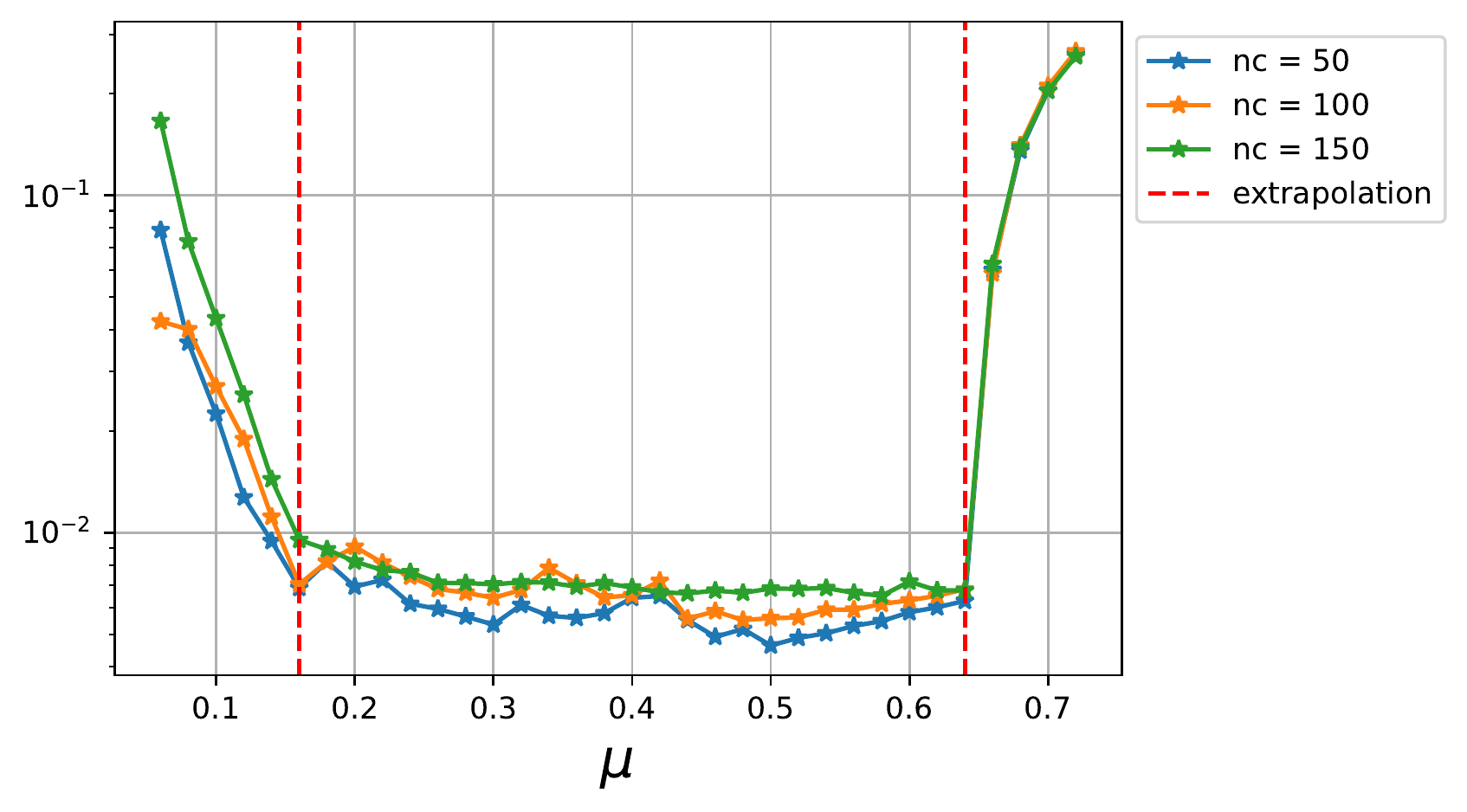}}
     \caption{$L^2$ relative and maximum errors of the hyper-decoder with Gappy-POD reconstruction of the full state for varying sizes of the sample mesh size ($n_c$). }\label{fig:nc_L2}
\end{figure}

 We analyze the influence of the sample mesh size, varying $n_c \in \{ 50, 100, 150 \}$ and using the stencil meshes in Figure \ref{fig:meshes}. In Figure \ref{fig:nc_L2} we report the $L^2$ and maximum relative errors for the FOM state reconstruction using our proposed hyper-decoder with LSPG and implicit time integration (BDF1 and BDF2). These experiments demonstrate our proposed method is accurate and in particular, integrating in time the ROM with BDF2 outperforms BDF1, with errors below 1\% across all the test parameters inside the training intervals and the different sample/stencil mesh sizes.
 
The red dashed line in Figure \ref{fig:nc_L2} delimits the interval of the training parameters. We observe that the errors increase at the extrapolation parameters (those outside the training interval). Nonetheless, in applications where a maximum relative error below 8\% would be acceptable, our proposed method with a sample-mesh size of  $n_c = 50$ and BDF2 time integration can be used with a trust region of $[0.8, 0.66]$.

Finally, previous works used hyper-reduction with GNAT-like methods to regularize the solution of the reduced order models with BDF1 time integration. Here we observe that a simple collocation strategy with more accurate time integration is enough to obtain accurate solutions of the ROMs. Notice moreover, that the overhead computational and memory costs of using BDF2 instead of BDF1 are proportional to the ROM state size, contrary to GNAT-like methods that require matrix multiplications of order $n_s$.  In practice, we find that using BDF1 and BDF2 for the ROM solutions leads to comparable running times.

 
\subsubsection{Extrapolation in time}

\begin{figure}[h!]
      \centering
        {\includegraphics[scale=0.7]{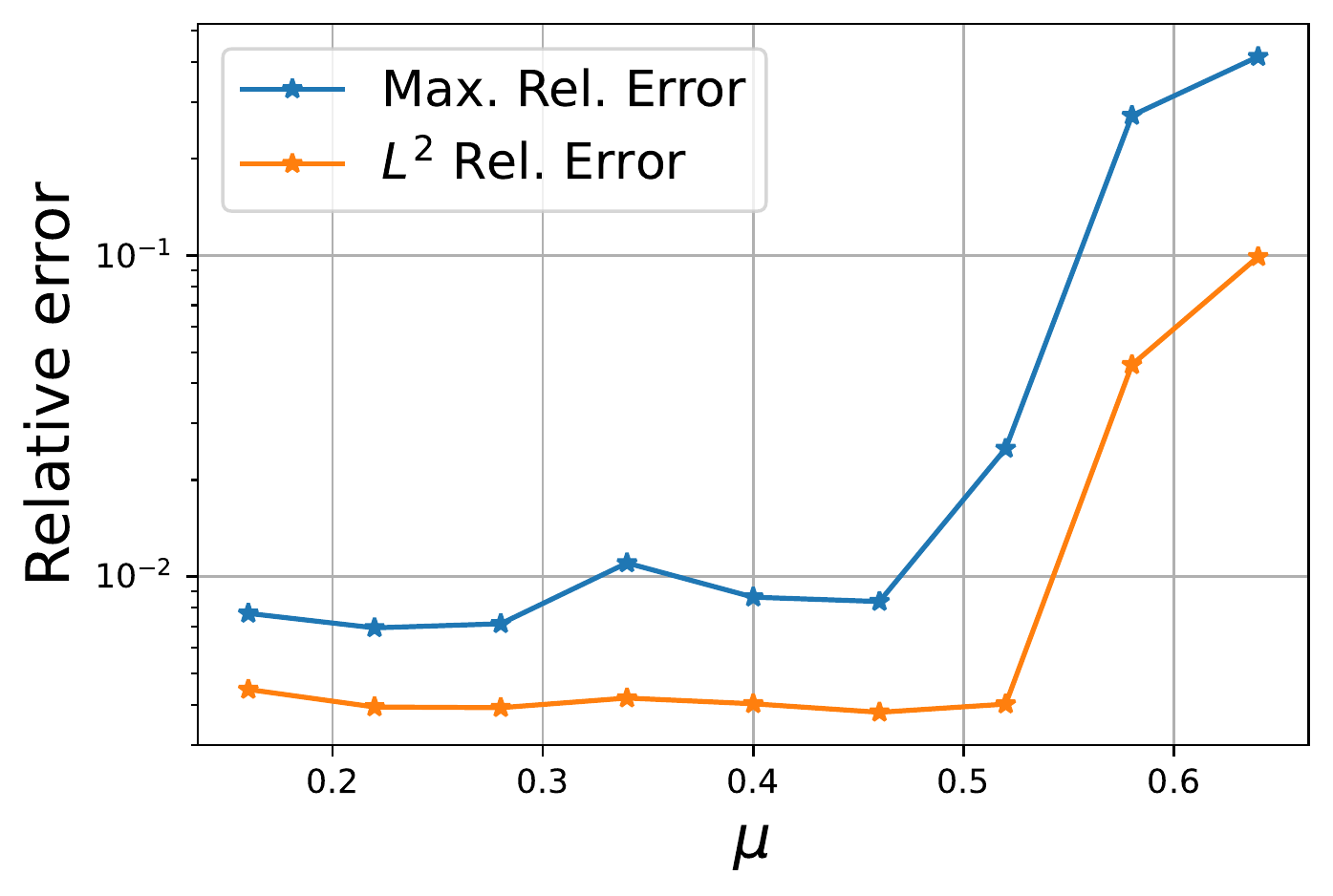}}
            \caption{Relative errors for the time extrapolation in the interval $[0,2.5]$. }\label{fig:extrap}
\end{figure}

We next study the ability of the proposed approach to extrapolate in time, that is to obtain accurate solutions even outside the time interval $[0, 2]$ that was used during training. We consider the 2d burgers equation in the time interval $[0, 2.5]$ with test parameters 
\[
	\mu \in \mathcal{D}_{\text{test}}' = \{ 0.16 0.22 0.28 0.34 0.4  0.46 0.52 0.58 0.64 \}.
\]
 We apply our proposed approach to estimate the FOMs in the extended time interval $[0, 2.5]$ and plot the $L^2$ and maximum relative errors in Figure \ref{fig:extrap}. We observe that a maximum relative error below 1.2\% is achieved for parameters $\mu \in [0.16, 0.46]$ and this trust region can be extended to $\mu \in [0.16, 0.52]$ accepting a maximum relative error below 3\%. These results suggest that the hyper-decoder has learned a meaningful representation of the reduced solution manifold, and our approach can (to some extent) be used to extrapolate in time. The decrease in accuracy for larger parameters is to be expected. In this range, the solutions travel more and the interactions with the boundaries become more important, making extrapolation harder.
 
\subsubsection{Effects of training size}

\begin{figure}[h!]
      \centering
        {\includegraphics[scale=0.7]{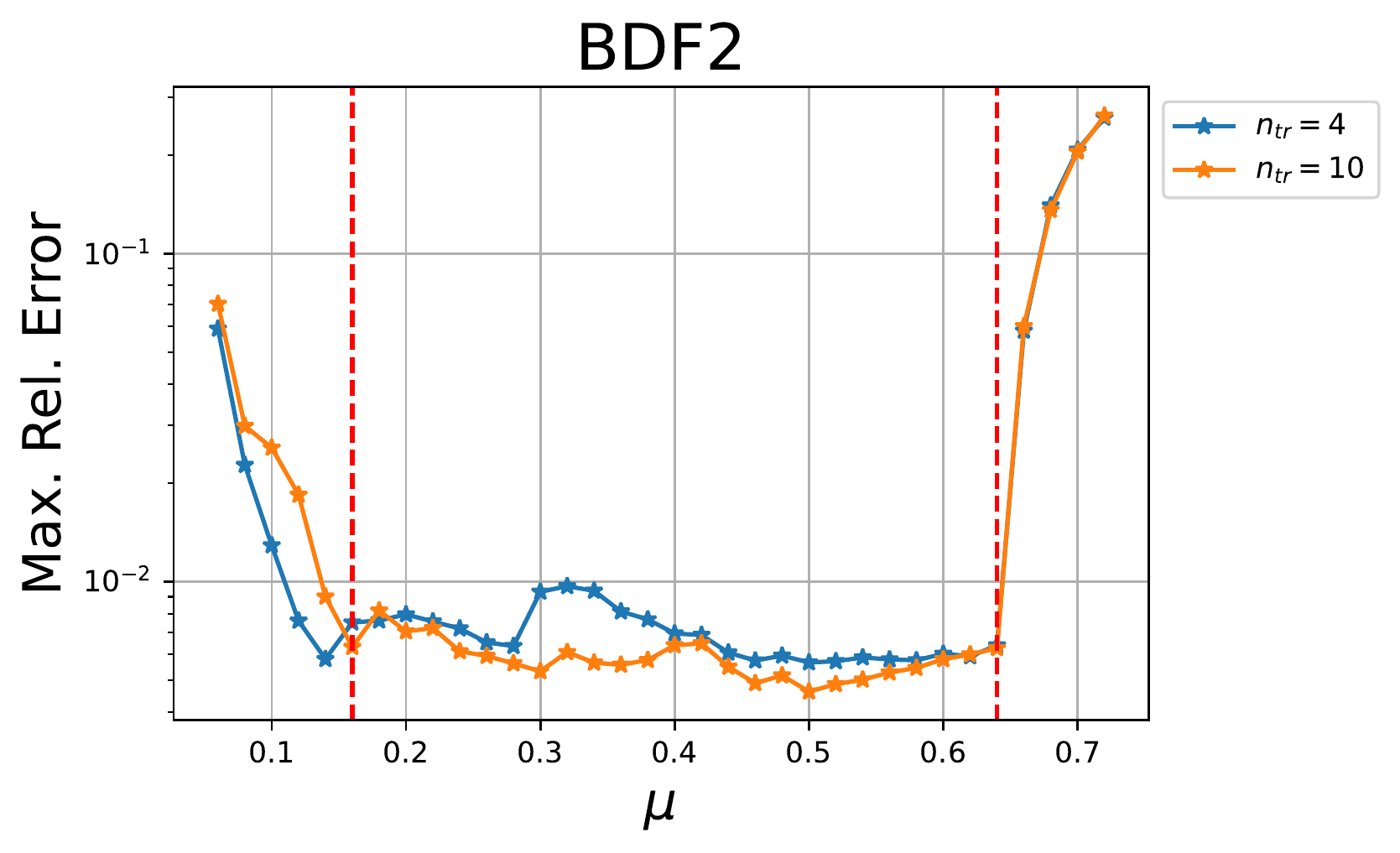}}
     \caption{Maximum errors of the hyper-decoder with Gappy-POD reconstruction of the full state for varying number $n_{tr}$ of parameters in the training set. }\label{fig:gen}
\end{figure}

In this section, we analyze the effects of the number of parameters used in the training set. We compare a model trained on $\mathcal{D}_\text{train}$ with $n_{tr} = 10$ number of parameters as in \eqref{eq:Dtrain}, and a model trained on  $\mathcal{D}_\text{train}'$ with $n_{tr} = 4$ number of parameters as in \eqref{eq:Dtrainp}). In both cases we use a sample mesh of size $n_c = 50$. The results summarized in Figure \ref{fig:gen} show that the errors of the two models are comparable and below the $1\%$ threshold in the training set interval. The model trained with $n_{tr} = 10$ perform slightly better than the model trained with $n_{tr} = 4$.

\section{Conclusions}\label{sec:conclusions}
    The use of nonlinear manifolds for projection-based reduced order models has been 
observed to yield improved accuracy relative to traditional linear subspace-reduced 
order models for a given ROM dimension, particularly for problems with slowly 
decaying Kolmogorov n-width such as advection-dominated ones.
 
These methods commonly use neural networks for manifold learning which often suffer
from high computational costs of training and evaluation relative to linear alternatives.
Recent works have proposed pruning \cite{choi} or network compression via teacher-student
training \cite{romor2022non} as approaches to reduce online evaluation costs.
However, these approaches do not address (and in general increase) offline training costs.

In this work, we develop and analyze a novel method that overcomes these disadvantages by 
training a neural network only on spatially subsampled versions of the high-fidelity
solution snapshots. This method coupled with collocation-based hyper-reduction and Gappy-
POD allows for efficient and accurate surrogate models in which neither the offline nor 
online cost scales with the high-fidelity model dimension.

We demonstrated the validity of our approach on a 2d Burgers problem.  These promising
initial results suggest that the use of hyper-reduced autoencoder architectures have
the potential to make nonlinear manifold projection-based reduced order models applicable
to large-scale problems which were previously not feasible due to impractical offline
training costs.

\cleardoublepage
\bibliographystyle{plain}
\bibliography{references}

\end{document}